\documentclass [a4paper,12pt]{article}

\begin{document}

\begin{center}
{\large \bf About the sign of $\Delta m_{12}^2$.}

\vskip 0.2in

Anatoly Kopylov\\ Institute of Nuclear Research of Russian Academy
of Sciences \\ 117312 Moscow, Prospect of 60th Anniversary of
October Revolution 7A
\end{center}

\footnotetext{Corresponding author: Kopylov A.V., Institute for
Nuclear Research of Russian Academy of Sciences, Prospect of 60th
Anniversary of October Revolution 7A, 117312, Moscow, Russia;
telephone (095)3340961, Fax (095)1352268, e-mail:
kopylov@al20.inr.troitsk.ru }

\begin{abstract}
The present status of the problem of identifying the sign of
$\Delta m_{12}^2$ is discussed in view of the development of new
methods of detection of solar neutrinos.
\end{abstract}

The results of solar neutrino experiments \cite{1}-\cite{5} plus
KamLAND \cite{6} give strong evidence that neutrinos are massive
particles and do oscillate \cite{7}. The important point is that
KamLAND experiment allows both signs for $\Delta m_{12}^2$ while
solar neutrino experiments restrict it only as a positive value.
If this is true then one can state that these experiments gave us
the very fundamental result that m$_2$ $>$ m$_1$. The impact of
this is so great that nearly in all reports and all papers by
describing the neutrino mass schemes only the case m$_2$ $>$ m$_1$
is analyzed. Should we accept that this is a firmly established
fact? Wise people say ``No'' \cite{8} because we still have no ``a
smoking gun evidence'' of this. We do have ``a smoking gun
evidence'' from SNO experiment \cite{5} that boron electron
neutrinos are attenuated by a factor of about three what is
expected from MSW effect for $\Delta m_{12}^2$ $>$ 0 \cite{9} and
this result is confirmed by a global analysis of all data \cite{7}
which prescribes also the attenuation factor of about two for low
energy neutrinos observed in a gallium experiment \cite{2},
\cite{3}. The agreement is impressive but still this is not enough
to make the final conclusion that m$_2$ is really heavier than
m$_1$. But what if not? Will this be a total collapse for the
solar neutrino experiments? This question was investigated in
\cite{10} and it was shown that it is possible to add to a
standard Hamiltonian a non standard part which can restore
agreement but the mixing angles can be different from what are
given now. The introduction of the new free parameters into a
theory is of course a costly affair but sometimes it works. What
about a mixing angle it appears to be that this point needs some
further clarification. And here the new experiments will be very
helpful. There are proposals to conduct the new experiments with
the reactor antineutrinos \cite{11} which can increase the
accuracy of measuring a mixing angle in comparison with KamLAND
although the sign of $\Delta m_{12}^2$ is a subtle thing. New
solar neutrino experiments \cite{12} will be also very helpful for
this. Especially interesting would be to make the measurements in
the intermediate energy (Be, pep and CNO neutrinos). The
measurement of the flux of beryllium and CNO neutrinos will be
very useful also for precise evaluation of the flux of
pp-neutrinos in the source (in the generation place) and the
tolerant uncertainties of these experiments are inversely
proportional to the weights of these sources in a total luminosity
of the Sun \cite{13}. All of this demonstrates that solar neutrino
research is still a viable source. If the mixing angles found in a
reactor antineutrino experiment and in the solar neutrino
experiments are different, this may be explained as:

1. the matter effect is not quite understood and not correctly
evaluated;

2. there's some New Physics (NSI etc.)

3. non CPT invariance;

4. something else happens in the Sun;

5. any of the solar neutrino experiments gives a wrong result.

In any case this will be a very important issue. The solar
neutrino problem is probably really solved but the suggested
solution may turn out to be very far from reality. Why not to
check it?

Acknowledgements.

I am grateful to Vadim Kuzmin who encouraged me to write this
notice.

\end{document}